\documentclass[%
 reprint,
 amsmath,amssymb,
 aps,
]{revtex4-2}
\usepackage{graphicx}% Include figure files
\usepackage{dcolumn}% Align table columns on decimal point
\usepackage{bm}% bold math
\usepackage{color}
\usepackage[normalem]{ulem}
\usepackage[dvipsnames]{xcolor}
\usepackage{comment}

\usepackage{mathtools,tikz}
\DeclareRobustCommand\sampleline[1]{%
  \tikz\draw[#1] (0,0) (0,\the\dimexpr\fontdimen22\textfont2\relax)
  -- (2em,\the\dimexpr\fontdimen22\textfont2\relax);%
}
\newcommand{\maj}{r_{\text{1}}}
\newcommand{\vdeform}{v_{1}}
\newcommand{\deformSTD}{\sigma_{v_1}}
\newcommand{\psd}{E_{11}^b}
\newcommand{\naturalfreq}{f_2}
\newcommand{\bubeddyvelocity}{\delta_D u}
\newcommand{\eddyvelocity}{\delta_l u}
\begin{document}
\title{Intermittency of bubble deformation in turbulence}
\author{Xu Xu}
\email{X. Xu and Y. Qi contributed equally to this work.}
\author{Yinghe Qi}
\email{X. Xu and Y. Qi contributed equally to this work.}
\author{Shijie Zhong}
\author{Shiyong Tan}
\author{Qianwen Wu}
\author{Rui Ni}
\email{Corresponding author. rui.ni@jhu.edu}
\affiliation{\textit{Department~of~Mechanical Engineering, Johns Hopkins University, 3400 N. Charles St., Baltimore, 21218, Maryland, USA}}

\begin{abstract}

The deformation of finite-sized bubbles in intense turbulence exhibits complex geometries beyond simple spheroids as the bubbles exchange energy with the surrounding eddies across a wide range of scales. This study investigates deformation via the velocity of the most stretched tip of the deformed bubble in 3D, as the tip extension results from the compression of the rest of the interface by surrounding eddies. The results show that the power spectrum based on the tip velocity exhibits a scaling akin to that of the Lagrangian statistics of fluid elements, but decays with a distinct timescale and magnitude modulated by the Weber number based on the bubble size. This indicates that the interfacial energy is primarily siphoned from eddies of similar sizes as the bubble. Moreover, the tip velocity appears much more intermittent than the velocity increment, and its distribution near the extreme tails can be explained by the proposed model that accounts for the fact that  small eddies with sufficient energy can contribute to extreme deformation. These findings provide a framework for understanding the energy transfer between deformable objects and multiscale eddies in intense turbulence.\color{black}
\color{black}
\end{abstract}

\maketitle
%%%%%%%% Variables %%%%%%
% Figure 2 change l=L/50 to l=2mm. Add legend for datapoints too
% Figure 3 add legend for lines
% figure 4 add legend for lines

% Major radius: $\maj$

% Major radius fluctuation: $\maj(t)-\langle \maj \rangle$

% Deformation veloicty: $\vdeform$

% STD of deformation velocity: $\deformSTD$

% Power spectrum of deformation velocity: $\psd$

% Bubble velocity: $\vbub$

% Average fluid velocity around bubble: $\vfluid$

% 2nd mode natural frequency: $\naturalfreq$

% 2nd mode natural frequency velocity: $\velnatural$

% Eddy velocity at bubble scale: $\bubeddyvelocity$

% Eddy velocity: $\eddyvelocity$

% eddy frequency at bubble scale: $\eddyfreq$

% eddy wave number at bubble scale: $\eddywavenumber$

% Model time scale: $\Tmodel$

% Model velocity scale: $\Umodel$

%%%%%%%%%%%%%%%%%%%%%%%%%%%%%%%%%%%%%%%%%%%%%%
%%%%%%%%%% Introduction %%%%%%%%%%%%%%%%%%%%%%
%%%%%%%%%%%%%%%%%%%%%%%%%%%%%%%%%%%%%%%%%%%%%%

\begin{figure} [t]
\centering
\includegraphics[clip,width=1\columnwidth]{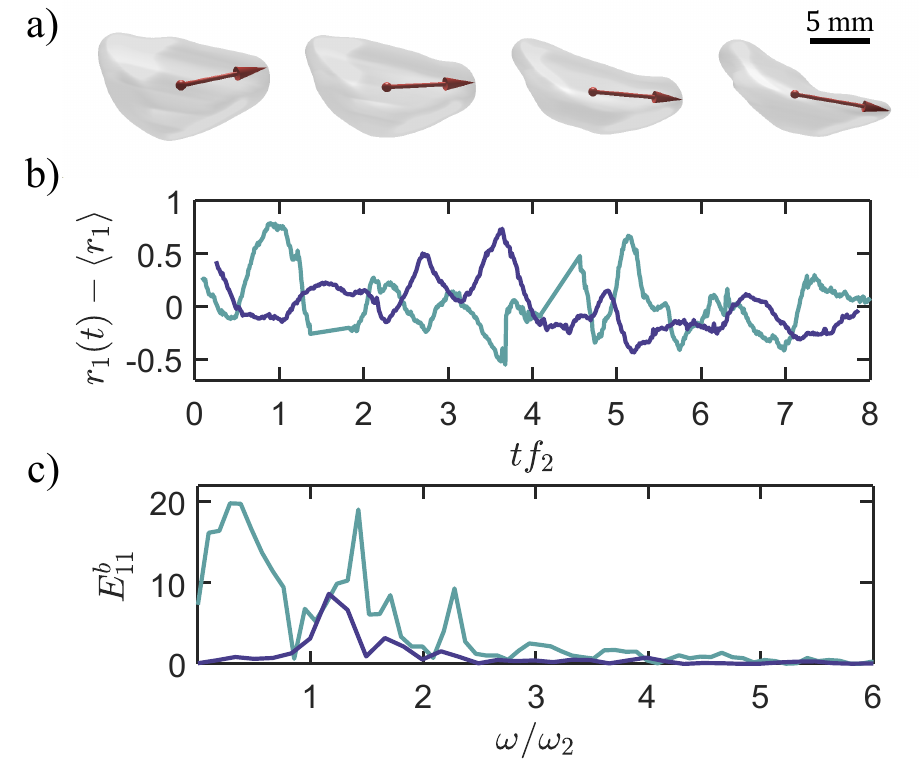}%

\caption{a) 3D reconstructed bubbles with dark red arrows indicating the semi-major axis $\maj$. b) Example time series of $\maj(t)- \langle \maj \rangle$ with normalized time $tf_2$. c) Power spectrum densities of $\vdeform$ with the normalized  angular frequency $\omega / \omega_2$. The two colored lines represent two example with mean bubble diameters of 8 mm (green) and 10 mm (blue).}
\label{fig:Fig1}
\end{figure}

%The deformation of bubbles driven by external flows serves as a prototypical example of understanding couplings between the two phases, from the early work by Lamb in deriving the natural oscillation of bubbles with small perturbation to the framework of. The 

As an enduring enigma within classical mechanics, turbulence has captivated countless minds because of the complex nonlinear interactions across a wide spectrum of scales. Adding a deformable and immiscible second phase in turbulence, e.g. bubbles or droplets, introduces a new set of complexities \cite{villermaux2007fragmentation, lohse2018bubble, elghobashi2019direct, mathai2020bubbly, ni2024deformation} and holds significant implications for applications such as two-phase heat \cite{albernaz2017droplet} and mass \cite{dodd2021analysis} transfer, air-sea interactions \cite{veron2015ocean}, and turbulence modulation \cite{crialesi2022modulation, perlekar2014spinodal, dodd2016interaction, rosti2019droplets}. The key questions arise as to (i) at which scale does the exchange between the turbulent kinetic energy and the interfacial energy occur \color{black}, (ii) how eddies of different sizes contribute to intermittent extreme deformation and breakup, and (iii) how to connect this deformation intermittency with the intermittency of the background turbulence.

These questions can be answered in a regime where the deformation is predominantly driven by turbulence instead of by buoyancy, which can be found in applications such as bubble-mediated drag reduction \cite{ceccio2010friction,murai2014frictional} and bubble fragmentation in breaking waves \cite{deane2002scale,gao2021bubble,chan2021turbulent}. Deformable bubbles are often in the inertial range of turbulence (with the diameter $D$ falls in $\eta \ll D\ll L$ with $\eta=(\nu^3/\epsilon)^{1/4}$ is the Kolmogorov scale and $L$ being the integral length scale). In the classical Kolmogorov--Hinze (KH) framework \cite{kolmogorov1949breakage, hinze1955fundamentals}, Kolmogorov stated that ``the breaking forces acting on them (drops or bubbles) due to the velocity differences, which are of the order of $u_D$'', in which $u_D$ represents the velocity increment at the bubble scale. Based on such a velocity scale, it was suggested that the deformation can be measured by the Weber number ($We_D$) defined as $We_D = \rho \bubeddyvelocity^2 D/\sigma$, where $\rho$ is liquid phase density  and $\sigma$ is surface tension coefficient. The velocity scale that drives the bubble deformation and breakup is assumed to correspond to the eddies with the size of the bubble, whose kinetic energy scales as $(\bubeddyvelocity)^2=C_2(\epsilon D)^{2/3}$ where the Kolmogorov constant $C_2$ is about 2 \cite{sreenivasan1995universality, ni2013kolmogorov}. For sufficiently large $We_D$, qualitatively, both experimental \cite{vejravzka2018experiments,lalanne2019model, qi2022fragmentation,qi2024timescale} and numerical \cite{perrard2021bubble,mangani2022influence} studies have shown deformed geometry clearly deviating from axisymmetric shape because the interfacial energy draws the kinetic energy from eddies of various scales instead of just the bubble diameter. In this letter, we aim at quantifying and understanding this deviation and illustrate its connection to turbulence intermittency.

The experiments were conducted in a vertical turbulent water tunnel \cite{masuk2019v} that can produce nearly homogeneous and isotropic turbulence (HIT). The turbulence was produced by a jet array which is located above the test section, facing down and randomly firing, with adjustable jet speeds allowing for a wide range of energy dissipation rates and Reynolds numbers. Turbulence decays as it moves away from the jet array \cite{tan2023scalings}. To quantify the background turbulence, the flow was seeded with tracer particles\cite{masuk2021simultaneous}, which were tracked in 3D by our open-sourced Lagrangian particle tracking method (openLPT) \cite{tan2019open, tan2020introducing}.  The fluctuation velocity $u'$ is around 0.2 m/s, and $L$ is about 60 mm. The Taylor microscale Reynolds number, i.e. Re$_\lambda=u'\lambda/\nu$, is roughly 435. The Taylor microscale $\lambda$ is defined as $\lambda=\sqrt{15\nu/\epsilon}u'$. The energy dissipation rate $\epsilon$ is around 0.16 m$^2/$s$^3$, from which the Kolmogorov length scale $\eta \approx 50$ $ \mu$m and timescale $\tau_\eta \approx 2.5$ ms can be determined. 

%The test section of the tunnel was built with an octagonal cross section to enable imaging from all sides. Six high speed cameras with the frame rate of 4000 frames per second were positioned across the tunnel perimeter to maximize the angles that they cover. 

Each camera has a dedicated LED light panel across the tunnel to cast dark silhouettes of deforming bubbles onto the camera's imaging plane. From these silhouettes, an in-house visual  hull algorithm was utilized to reconstruct the 3D geometries of the bubbles \cite{masuk2019robust}, which are shown in Fig. \ref{fig:Fig1} (a). The spherical-equivalent diameter ($D$) of this example bubble is roughly 8 mm. The diameter of the bubbles used in this work ranges from $20\eta$ to $200\eta$ (1--10 mm), which is within the inertial subrange of the ambient turbulence. 
%
%\color{black}
%The problem obeys the Navier-Stokes equations with an additional term that denotes the force exerted by the surface tension
%\begin{equation}
%    \rho\left(\partial_t u_i+u_j \partial_j u_i\right)=-\partial_i p+\partial_i\left[\mu\left(\partial_i u_j+\partial_j u_i\right)\right]+f_i^\sigma
%\end{equation}
%
%where $u_i$ is the velocity in the $i$ th direction, $p$ is the pressure, and $\rho$ and $\mu$ are the local density and viscosity. The forcing term $f_i^\sigma=\sigma \delta(s)\xi n_i$ represents the surface tension force that acts only at the interface, where $\sigma$ is the surface tension, $\xi$ is the local interface curvature, $n_i$ is the $i$ th component of the unit vector that is normal to the interface and directed towards the interior of the bubble, and $\delta(s)$ denotes the Dirac $\delta$-function that is needed to impose $f_i^\sigma$ only at the interface position with the coordinate $s=0$ marking the interface location. This equation can be further converted to the energy equation by multiplying both sides with $u_i$. The term of $f_i^\sigma$ becomes $\mathcal{S}_\sigma=f_i^\sigma u_i$, which represents the work of the surface tension force. Since $f_i^\sigma$ is nonzero only at the interface, $u_i$ becomes the velocity of the interface. 

The work done by interface to the surrounding liquid and vice versa is the product of the surface tension coefficient, the interface curvature, and the interface velocity. Out of all the interfacial points, the most extruded point of the bubble interface (the tip) has the largest curvature and interfacial velocity, which indicates that its contribution to the work by surface tension is the greatest. In addition, for complex deformation, the local compressions induced by pressure fluctuations over different parts of a bubble driven by eddies of various scales collectively result in an extension along the longest axis. As a result, the tip velocity can be indicative of the multiscale exchange between kinetic and interfacial energy. Note that this velocity carries information about the intermittent events in turbulence which will be filtered out if using the velocity of the semi-major axis after fitting the geometry with an ellipsoid. From the reconstructed shapes, $\maj$ is defined as the vector extending from the center of mass to the vertex that is furthest away from the center, which is shown as dark red arrow in Fig. \ref{fig:Fig1} (a). The reconstruction accuracy for the semi-major axis is the highest due to the ease of observation of the most extruded tip by multiple cameras without being obstructed. 

%Furthermore, in cases of irregular deformation, uncertainties in determining the center of mass have the least impact on the length of the semi-major axis compared to other axes.

The time series of the semi-major axis fluctuation $\maj(t)-\langle \maj \rangle$ for two example bubbles are shown in Fig. \ref{fig:Fig1} (b), and the time is normalized with the Lamb 2$^{\text{nd}}$ mode natural frequency $\naturalfreq=  \sqrt{96\sigma/(\rho D^3)}/2\pi$ \cite{horace1932hydrodynamics}, which was suggested to be the characteristic timescale for bubble deformation in turbulence \cite{risso1998oscillations}. $\maj(t)$ exhibits complex fluctuations, covering a wide range of scales in time. To further illustrate the Lagrangian dynamics, the power spectrum density (PSD) of the corresponding velocity $\vdeform=d\maj(t)/dt$ is shown in Fig. \ref{fig:Fig1} (c), and the angular frequency is normalized as $\omega_2= 2\pi \naturalfreq$.   For the examples shown, the majority of energy that measures the bubble deformation does not always peak at $\omega = \omega_2$, suggesting that the bubble deformation is dominated by timescales of surrounding eddies rather than its natural oscillation. 

\begin{figure} [bth!]
\includegraphics[clip,width=1\columnwidth]{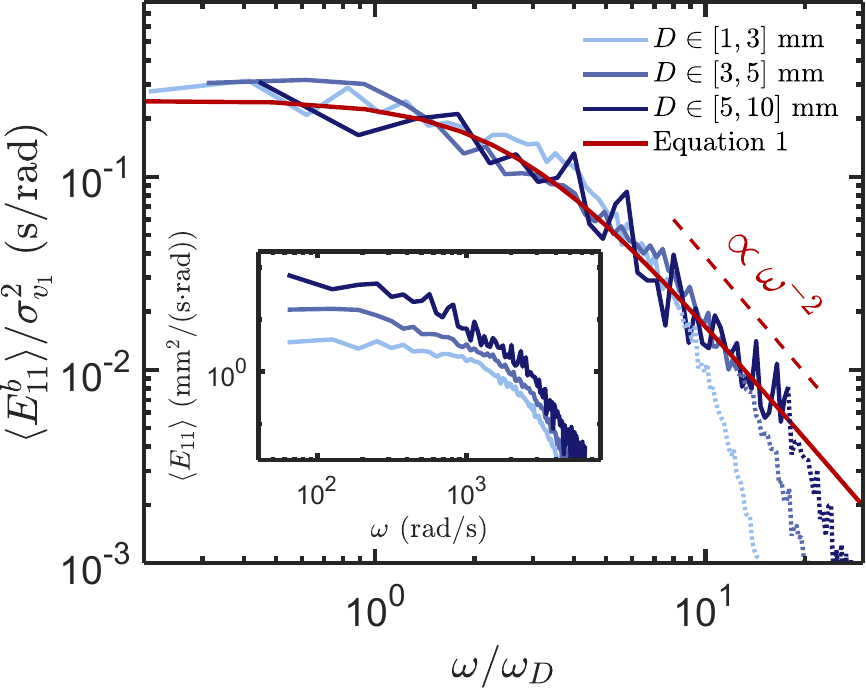}%
\caption{The normalized power spectrum densities of the deformation velocity for different bubble sizes. The red solid line is equation \ref{eq1} normalized with $\deformSTD^2$. The solid blue lines switch to dotted lines with the same color at $\tau_\eta$. The inset shows the power spectrum densities for the three bubble size bins. }
\label{fig:Fig2}
\end{figure}

To further illustrate how surface tension and external turbulence affect the deformation dynamics, the PSD was calculated for all bubble trajectories similar to the sample cases in Fig. \ref{fig:Fig1}. Since individual bubble trajectories are different in length, the PSD of $\vdeform$ was calculated first for each bubble trajectory that is at least longer than one turnover time of eddies with the size of the bubble. The PSDs of individual bubble trajectories are then projected onto a commonly discretized frequency range. In each frequency bin, we conducted weighted average where the lengths of the trajectories are used as the weight.

%the PSDs are evaluated at different frequencies determined by the length of the individual bubble track length. To collectively study the PSDs, we ensemble average the individual PSD with a uniformly discretized frequency space from 10 to 1000 Hz 
%with the individual bubble track length as its corresponding weight. \color{black}Two additional datasets collected in similar flow conditions are included in the PSD calculation:  turbulence induced by the head-on collision of two vortex rings \cite{qi2022fragmentation}, and HIT generated in a new vertical water tank \cite{qi2024timescale}. The resulting PSD of $\vdeform$ for different bubble sizes are shown in the inset of Fig. \ref{fig:Fig3} \color{blue}, where 181, 221, and 70 bubble tracks are utilized for the three bubble size bins from small to large. \color{black}

Two additional datasets collected in similar flow conditions are included in the PSD calculation:  turbulence induced by the head-on collision of two vortex rings \cite{qi2022fragmentation}, and HIT generated in a new vertical water tank \cite{qi2024timescale}. In the inset of Fig. \ref{fig:Fig2}, the  averaged PSD systematically shifts up as the bubble size increases, suggesting that larger bubbles are more deformable and exhibiting stronger deformation. For the bubble sizes examined, PSDs show only plateaus at low frequencies with no clear peaks after averaging as the dominant frequency for each trajectory varies over a range. The plateau indicates the range of frequencies that bubbles absorb kinetic energy from surrounding turbulent eddies. This plateau transitions to rapid decay at higher frequencies, with the transition frequency seemingly decreasing as bubble sizes increase. By normalizing the horizontal axis with the eddy turnover angular frequency $\omega_D=2\pi\sqrt{2}\epsilon^{1/3}D^{-2/3}$ 
 and the vertical axis with $\deformSTD^2$, the PSDs for different bubble sizes collapse up to the Kolmogorov frequency $2\pi/\tau_\eta$, which is illustrated in the main panel of Fig. \ref{fig:Fig2}.

The decay of the normalized PSD above the Kolmogorov scale follows a power law scaling $\psd(\omega)\propto \omega^{-2}$. It is interesting to find that it resembles the scaling observed in the Lagrangian energy spectrum of the single-phase turbulence $E^{L}_{11}$, which is defined as the Fourier transform of the Lagrangian velocity autocorrelation function $\langle u(t)u(t+\tau) \rangle_t/ \langle u^2 \rangle$. For the time delay $\tau$ in the inertial sub-range ($\tau_\eta\ll\tau\ll T_L$ with $\tau_\eta$ being the Kolmogorov timescale and $T_L$ being the integral timescale), it has been shown that $E^{L}_{11}$ follows a simple scaling law of $\epsilon \omega^{-2}$, which can be derived from the dimensional analysis \cite{mordant2001measurement}. This similar scaling points out a possible connection between the deformation of the semi-major axis of a bubble to the deformation of a Lagrangian fluid element. 

\begin{figure} [bth]
\includegraphics[clip,width=1\columnwidth]{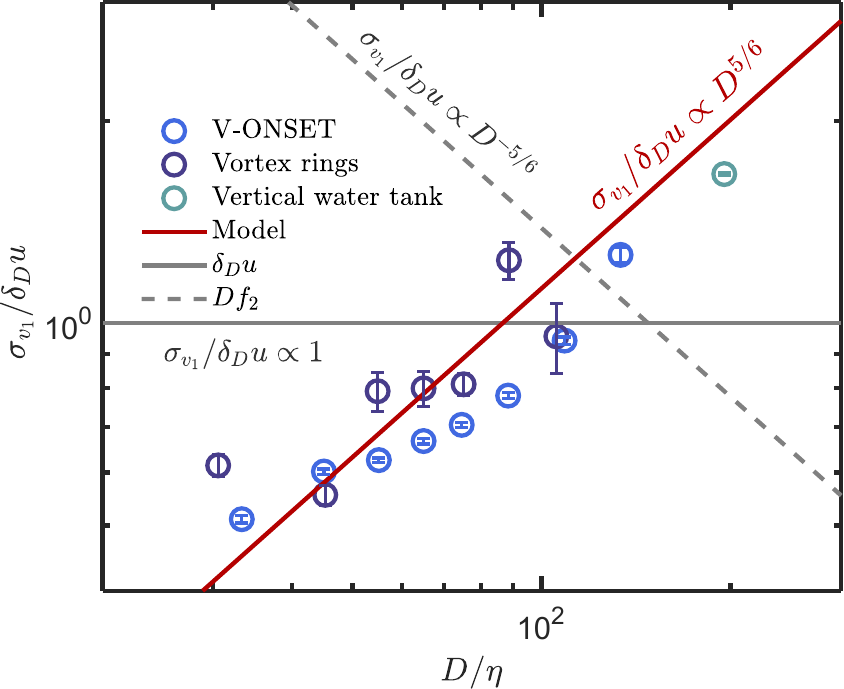}%

\caption{Bubble deformation velocity normalized by the bubble-sized eddy velocity $\bubeddyvelocity$ as a function of the bubble diameter $D$ normalized by the Kolmogorov scale $\eta$. The three lines indicate three deformation velocity models.}
\label{fig:Fig3}
\end{figure} 

In single-phase turbulence, the Lagrangian velocity of a fluid element decorrelates at the integral timescale $\tau=T_L$, and its power spectrum can be expressed as $E^L_{11} = (u'^2T_L)/[1+(\omega T_L)^2]$ \cite{mordant2001measurement}. Under conditions of negligible surface tension and perfectly matched density and viscosity, the interface of a deformable bubble (or equivalently a finite-sized fluid element) should indeed follow the Lagrangian trajectory and their velocity would match. However, the presence of surface tension and differences in densities and viscosities between inner and outer fluids lead to the velocity difference, i.e. not all the kinetic energy available from the surrounding turbulence $u'^2$ can be transferred into the interfacial energy. Thus, by modifying $E^L_{11}$ in a proper way, one could model the power spectrum of the interfacial velocity $\psd$ as   follows

\begin{equation}
\psd (\omega) =  \frac{2 \deformSTD^2/\omega_b}{\pi [1+(\omega/\omega_b)^2]} 
\label{eq1}
\end{equation}
where $\omega_b$ represents the angular frequency for energy injection from turbulence into the interface and signifies the duration during which the interfacial velocity remains correlated. The variance of deformation velocity, $\deformSTD^2$, quantifies the interfacial kinetic energy associated with bubbles of size $D$. It can be seen from Eq. \ref{eq1} that $\omega_b$ determines the transition from a plateau to a $\omega^{-2}$ scaling at higher  frequencies  \color{black}. By setting  $\omega_b=\omega_D/3$ with a fitting prefactor of 1/3, a good agreement between Eq. \ref{eq1} and the experimental data is achieved, as illustrated in Fig. \ref{fig:Fig2}. This agreement suggests that the transition timescale is governed by the bubble-sized eddies with a frequency close to $\omega_D$, rather than the second-mode natural oscillation frequency $\omega_2$. 

The overall energy of the PSD can be indicated by the integration of $E_{11}^b$  over $\omega$ which yields  $\deformSTD^2$. Fig. \ref{fig:Fig3} shows the dependence of $\deformSTD$ on $D/\eta$ using the same three datasets as aforementioned in Fig. \ref{fig:Fig2} \cite{qi2022fragmentation, qi2024timescale}. A consistent trend with size $D$ among all three datasets is observed.  

To probe the relationship between $\deformSTD$ and the bubble size, two extreme scenarios should be considered: (i) The deformation is entirely driven by external forces so that the surface tension effect is negligible, or (ii) The deformation is primarily dominated by the natural oscillation when external perturbations are weak. In the first scenario, $\deformSTD$ scales linearly with the eddy velocity, i.e. $\deformSTD\sim\bubeddyvelocity$. As a result, $\sigma_{v_1}/\delta_D u$ should show no dependence on $D/\eta$, as illustrated by the horizontal line in Fig. \ref{fig:Fig3}.  For the second case in which the natural oscillation dominates, the interfacial velocity can be approximated by the characteristic velocity associated with the natural frequency, $\deformSTD/\bubeddyvelocity \propto D\naturalfreq/\bubeddyvelocity = \pi^{-1}  \epsilon^{-1/3} \sqrt{12\sigma/\rho} D^{-5/6}$, resulting in a power law scaling of $D^{-5/6}$, as depicted by the dashed line. However, it is evident that neither of these scaling models adequately explains the measured interfacial velocity. 

%The interfacial velocity exhibits a scaling with $D$ at a higher exponent than what is suggested by the two proposed limits.

%The dependence of $\deformSTD$ on bubble size can be modelled by considering two limits: either the deformation is dominated completely by the external forces or is driven by its natural oscillation assuming that the external perturbation is weak. If $\deformSTD$ scales linearly with the eddy velocity, i.e. $\bubeddyvelocity$, the result should follow the solid horizontal line in Fig. \ref{fig:Fig4} with no dependence on $D/\eta$ after $\deformSTD$ being normalized by the eddy velocity. Alternatively, if the interfacial velocity scales with the characteristic velocity associated with the natural frequency, $\deformSTD/\bubeddyvelocity$ would scale with $D\naturalfreq/ \bubeddyvelocity =1/(2\sqrt{2}\pi \epsilon ^{-1/3}) \sqrt{96\sigma/\rho} D^{-5/6}$, which follows a negative scaling with $D$ after normalization, as shown by the dashed line. Clearly, neither of these two scales explains the measured interfacial velocity. 

\begin{figure} [bth!]
\includegraphics[clip,width=0.95\columnwidth]{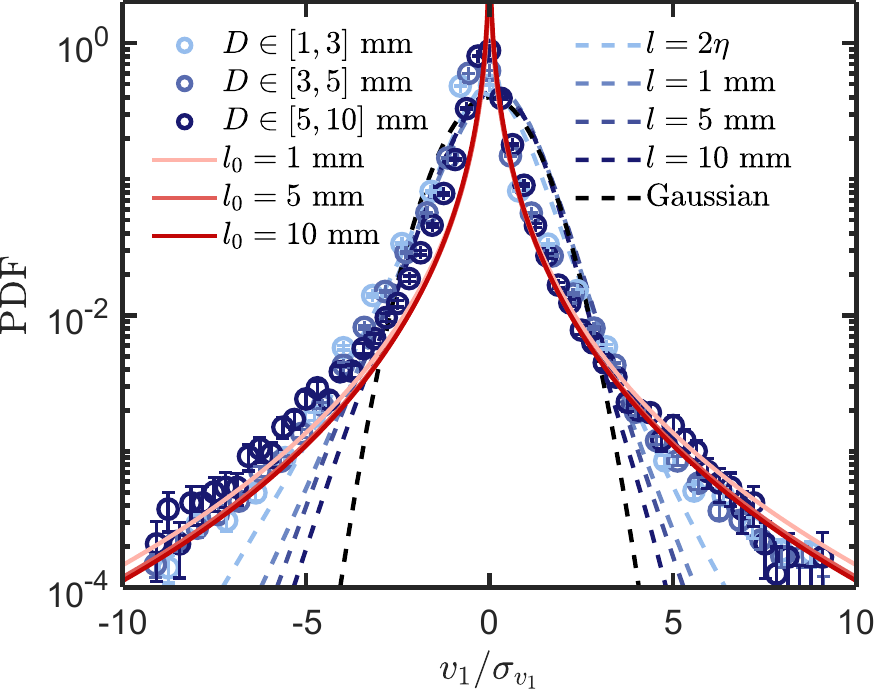}%

\caption{The probability density function of the bubble deformation velocity $\vdeform$ normalized by their respective standard deviations $\sigma_{v_1}$. In addition, the black dashed line indicates the standardized normal distribution. The blue dashed lines represent the modelled PDF of the longitudinal velocity increment, following the work by \cite{chevillard2006unified}, with the separation distance $l$ matching that of the bubble sizes. The red lines are from our model, where $l_0$ denotes the largest length scale. }
\label{fig:Fig4}
\end{figure}

\color{black} Note that small bubbles with large surface tension can barely deform, making it difficult for their interfacial velocity to match the eddy velocity nearby. This physical picture contradicts the first scenario, i.e. $\deformSTD \sim \bubeddyvelocity$. Therefore, the model can be improved by adding the modulation by surface tension, measured by the Weber number, to the framework. For Weber numbers on the order of one  ($We_D\sim\mathcal{O}(1)$) as explored in this letter, this modulation of the deformation energy, in the leading order, can be assumed to be linear, specifically $\deformSTD^2\sim We_D(\bubeddyvelocity)^2$. Rearranging this equation leads to $\deformSTD/\bubeddyvelocity\sim\sqrt{We_D} \propto D^{5/6}$ as represented by the red solid line in Fig. \ref{fig:Fig4}. The predicted scaling aligns well with the experimental results in all the three datasets, covering roughly a decade of the bubble size. This finding suggests that turbulence-driven bubble deformation is primarily induced by nearby eddies close to the bubble size but modulated by the Weber number.

%Moreover, the energy decay over frequency mirrors that of the Lagrangian power spectrum in single-phase turbulence, despite the complexities introduced by mismatched density and surface tension. 

To further investigate the deformation velocity, Fig. \ref{fig:Fig4} shows the full probability density functions (PDFs) of $\vdeform$ for bubbles with different sizes, from $D=1$ to 10 mm, with about $10^5$ data points used for each size bin. The PDFs are normalized by the respective $\deformSTD$. A comparison between the PDF and the standardized normal distribution (black dashed line) highlights the intermittent nature of bubble deformation. Notably, in the vicinity of the peak, the PDF exhibits asymmetry around zero, indicating a slightly elevated probability of encountering events featuring negative $\vdeform$ due to the preferential roles of surface tension that leads to tip retraction.

%Particularly, $\eddyvelocity$ exhibits similar skewness towards negative velocity \citep{mordant2001measurement, chevillard2006unified}, postulating that the skewness observed in $\vdeform$ comes from that of $\eddyvelocity$ as turbulence imprints on the bubble deformation. 

In the limit of negligible surface tension and perfect density match between the two phases, $\vdeform$ is anticipated to mirror the Eulerian velocity increment of single phase turbulence, denoted as $\eddyvelocity$, between two points separated by a distance $l$ equating the semi-major axis of the deformed bubble. In Fig. \ref{fig:Fig4}, the blue dashed lines represent the PDFs of the velocity increment $\mathcal{P}(\eddyvelocity)$ based on the model by Chevillard et al. \cite{chevillard2006unified}, encompassing scales from $l= 2\eta$ to $10$ mm, corresponding to our bubble size range. A significant contrast can be observed between $\mathcal{P}(\eddyvelocity)$ and $\mathcal{P}(\vdeform)$. 

%Particularly, $\mathcal{P}(\vdeform)$ exhibits stronger intermittency than $\mathcal{P}(\eddyvelocity)$ overall, but it does not show increased intermittency as $\mathcal{P}(\eddyvelocity)$ does for smaller $l$. %The overall difference may come from the mismatched density and viscosity that leads to the deviation of the deformation velocity from the velocity increment. The absence of varying intermittency of $\vdeform$, on the other hand, could be resulted from the surface tension effect, which becomes stronger for smaller bubbles and resist the rarer but more extreme deformation events. 

The key difference between $\mathcal{P}(v_1)$ and $\mathcal{P}(\delta_l u)$ is that bubble deformation can be contributed by eddies with more than one size \cite{qi2022fragmentation}. Assuming the bubble encounters an eddy with a size $l<D$ in turbulence, the bubble deformation velocity $v_1^l$ induced by this eddy can be modeled using $v_1^l\sim We_l^{1/2}\delta_l u$ by considering the modulation of the Weber number following the previous discussion. Here, $\delta_l u$ denotes the velocity scale of the sub-bubble-scale eddy and $We_l=\rho\delta_l u^2 l/\sigma$ is the corresponding Weber number \color{black}that measures the sub-bubble-scale local deformation\color{black}. Substituting $We_l$ in to $v_1^l$ leads to $v_1^l\color{black}\propto\color{black}\delta_l u^2 l^{1/2}$. This indicates that the PDF of $v_1^l$ is more intermittent than  $\mathcal{P}(\delta_l u)$ under the modulation of $We_l$. Note that $\mathcal{P}(\delta_l u)$ can be fitted using stretched exponentials following $\mathcal{P}(\delta_l u)=\sum_{k=0}^n C_k^n 2^{-n}(2\pi\sigma_{k;n}^2)^{-1/2}\text{exp}\left[-\delta_l u^2/(2\sigma_{k;n}^2)\right]$, where $n=\log_2(L_0/l)$ by assuming binary cascade in turbulence and $\sigma_{k;n}=\langle\Delta u_0^2\rangle^{1/2}M^{k/3}(1-M)^{(n-k)/3}$ with $\Delta u_0$ being the typical velocity increment at the macroscopic scale and $M=0.3$ being a fitting parameter \citep{kailasnath1992probability}.  Then the PDF of  $v_1^l$, $\mathcal{P}(v_1^l)$ can be derived following $\mathcal{P}(v_1^l)=\mathcal{P}(\delta_l u)\text{d}(\delta_l u)/\text{d}v_1^l \color{black}\propto\color{black} \mathcal{P}(\delta_l u)\delta_l u^{-1} l^{-1/2}$. $\mathcal{P}(v_1^l)$ is the probability distribution of bubble deformation velocity when the bubble encounters the eddies only with size $l$. In practice, eddies of various sizes exist simultaneously and small eddies are more abundant than large ones. To incorporate this effect on the PDF of the deformation velocity, we introduce the bubble-eddy collision frequency $\omega_c \propto \color{black}l^{-11/3}$ \cite{qi2022fragmentation,qi2024timescale,luo1996theoretical} and then calculate the weighted average of $\mathcal{P}(v_1^l)$ caused by eddies of different length scales.  This method gives the PDF of $v_1$, i.e., $\mathcal{P}(v_1)=\int_\eta^D\mathcal{P}(v_1^l)\omega_c\text{d}l/\int_\eta^D\omega_c\text{d}l$. In this equation, the contribution from all the sub-bubble-scale eddies, with sizes from $\eta$ to the bubble size $D$ is incorporated.
%the notion of local Weber number does not apply when the eddy size is the same as the bubble size. \textcolor{blue}{I don't think this is the reason why the model does not work for small deformation velocity. When the deformation velocity is small, the Weber number is weak and the surface tension effect is significant. So 

The current model should \color{black}only be effective at predicting $\mathcal{P}(v_1)$ near the tails when $v_1$ is large, not when $v_1$ is close to zero\color{black}, where the predicted PDF diverges and the interface velocity is not anticipated to follow the fluid velocity increment. As a result, the predicted $\mathcal{P}(v_1)$, as shown by the red lines in  Fig. \ref{fig:Fig4}, is shifted vertically to facilitate a comparison of the PDF tails. A good agreement with the experimental data is evident for various bubble sizes. In particular, the model captures the strong intermittency of the bubble deformation velocity marked by the longer tails compared with $\mathcal{P}(\delta_l u)$. This stronger intermittency originates from two aspects: (i) the contribution from Kolmogorov-scale eddies whose velocity distribution shows significant intermittency and (ii) the modulation by the local Weber number $We_l$ as mentioned above.  
%at around zero because . This is likely due to the fact that the small deformation velocity due to the capillary waves on the bubble interface is not considered, espeically when the Weber number is negligible. (Need a better reason?)

\color{black}

In summary, bubbles primarily absorb energy from eddies of comparable size to induce deformation; however, they can exhibit extreme deformation intermittently as interfaces are also affected by intense small-scale eddies. Surprisingly, bubble deformation is more intermittent than the surrounding turbulence because the modulation of deformation by local surface tension effectively results in a biased selection of extreme events in bubble deformation. These findings provide new insights into deformation and breakup dynamics in intense turbulence, extending beyond simple linear frameworks and offering new insights into bubble behavior in turbulent environments.

\color{black}

%Despite the highly nonlinear interactions between the two phases and large perturbations introduced by surrounding eddies across a wide range of scales, the resulting deformation velocity exhibits similar behavior as the Lagrangian velocity of fluid elements, but modulated linearly by the Weber number. This finding indicates a leading order balance between turbulent inertial forcing and surface tension in governing the interfacial dynamics for the  Weber number in the order unity. Looking forward, incorporating effects from viscosity contrast and internal circulation would enable more comprehensive modeling across a broader parameter space. More importantly, the improved understanding of the multiscale two-phase interactions could help advance the modelling and parameterization of momentum, mass, and heat transfer across a deformable interface. 

\begin{acknowledgments}
This work is supported by the Office of Naval Research  (Grant \#: N00014-21-1-2083). 
\end{acknowledgments}

X. Xu and Y. Qi contributed equally to this work.

\bibliographystyle{apsrev4-1} % Tell bibtex which bibliography style to use
%\bibliography{XuCite} % Tell bibtex which .bib file to use (this one is some example file in TexLive's file tree)

\end{document}